\documentclass[journal]{IEEEtran}
\usepackage{cite}
\usepackage{amsmath,amssymb,amsfonts}
\usepackage{algorithmic}
\usepackage{graphicx}
\usepackage{multirow}
\usepackage{textcomp}
\usepackage{url}
\usepackage{caption}
\ifCLASSINFOpdf
\else
\fi
\begin{document}
%
\title{A Review on Serious Games for Exercise Rehabilitation}
%
%
%

\author{Huansheng~Ning,~\IEEEmembership{Senior~Member,~IEEE,}
        Zhenyu~Wang, Rongyang~Li, Yudong~Zhang,~\IEEEmembership{Senior~Member,~IEEE,} Lingfeng~Mao
\thanks{Huansheng~Ning, Zhenyu~Wang, Rongyang~Li, Lingfeng~Mao are with the School of Computer and Communication Engineering, University of Science and Technology Beijing, 100083, Beijing, China (email: ninghuansheng@ustb.edu.cn).}
\thanks{Huansheng~Ning is also with Beijing Engineering Research Center for Cyberspace Data Analysis and Applications, 100083, Beijing, China.}
\thanks{Yu-Dong Zhang is with the School of Informatics, University of Leicester, Leicester, LE1 7RH, UK, e-mail: yudongzhang@ieee.org}}


%
%

\markboth{Journal of \LaTeX\ Class Files,~Vol.~14, No.~8, August~2015}%
{Shell \MakeLowercase{\textit{et al.}}: Bare Demo of IEEEtran.cls for IEEE Journals}
%



\maketitle

\begin{abstract}
Disability is an important factor affecting today’s society. At the same time, more and more sub-healthy people are sick due to reduced body functions and cognitive functions. Exercise rehabilitation is a kind of physical therapy, which can recover the motor ability, cognitive ability, and mental state of them through exercise. But the traditional exercise rehabilitation has some drawbacks so that people who need exercise rehabilitation cannot stick to it. Therefore, many researchers improved the drawbacks of traditional exercise rehabilitation by serious games for exercise rehabilitation. Although there were abundant achievements in the games, its relevant technologies and representative games are not be summarized systematically. To fill this gap, we introduced the significance of the convergence of exercise rehabilitation and serious games. Then, our paper sorted out the development of the games based on interaction mode between games and players. Besides, we analyzed the characteristics of different user groups and the specific functions of the games corresponding to them, and gave our classification based on this. Based on the classification, we reviewed related studies of the games in the past decade years and gave some suggestions on game design and development. Finally, we proposed serval research directions worth studying about the games’ technology development, functional design and social popularization.
\end{abstract}


%
\IEEEpeerreviewmaketitle

\section{Introduction}
%
%
%
%
\IEEEPARstart{A}{} survey by the World Health Organization (WHO) shows that the truly healthy people, the diseased people, the sub-healthy people account for about 5\%, 20\%, 75\% respectively\cite{Subhealth1}. People in a sub-health state often feel fatigued, muscle and joint pain, depression, inability to concentrate, memory loss, social difficulties, etc \cite{Subhealth2}. These phenomena affect a person's normal life to varying degrees. Another survey by WHO in 2014 showed that around 15\% of the world's population (more than 1 billion people) live in some form of disability, and 2-4\% of them have serious difficulties at work \cite{Disability}. By 2050, this figure will double to 2 billion \cite{Disability}. Disability is a compound factor that affects many aspects of a person's life. Compared with people without disabilities, people with disabilities have poorer health, fewer opportunities for education and work, and a higher probability of poverty in life \cite{Disability}. Rehabilitation is very important for people with sub-health status and disabled patients. Exercise rehabilitation is a way to recover sub-healthy people and patients through exercise. It is a commonly used method in the rehabilitation process. However, the traditional exercise rehabilitation process is single and repetitive, so it is often difficult for people who need exercise rehabilitation to maintain interest in it, which lead to bad rehabilitation effect. So, researchers try to find a new way to achieve exercise rehabilitation.
\par Serious games are software that combines non-entertainment purposes with the structure of games \cite{djaouti2011classifying}. These games often provide professional training and simulation and have a clear educational purpose. Serious games have interestingness, which can make patients have certain compliance with the treatment process and obtain good treatment effects. Hence, more and more researchers have used serious games for exercise rehabilitation (SGER) to treat people who need exercise rehabilitation. \cite{bonnechere2016use}, \cite{nawaz2016usability} have confirmed the effectiveness of SGER, which is more interesting than traditional exercise rehabilitation.
\par There have been precedents of using games for exercise rehabilitation a long time ago. These games required players to directly interact with game props. Such as building block games can help people who need exercise rehabilitation improve limb motor ability and cognitive ability, and rope skipping can help them improve lower limb motor ability and balance ability. Later, with the rise of video games, there have been many SGERs, which are controlled by the controller. Reinkensmeyer et al. \cite{1031978} provided an upper limb rehabilitation program after stroke, which was called Java therapy. In this therapy, players need to control a joystick (named Logitech Wingman Forcefeedback Pro) to complete the rehabilitation game "Breakout Therapy". Fruites \cite{7800105} was a game controlled by Kinect, which was used for the rehabilitation of children with autism. Although there were many related studies on SGER, no researchers have reviewed the development of SGER so far. In order to fill this gap and help researchers better understand SGER, we reviewed the development of SGER from the perspective of game interaction method based on hardware. 
\par In recent years, a huge number of studies on SGER have promoted the development of SGER. Researchers reviewed these existing SGERs from different angles. Such as, some researchers have specifically studied SGER for treating post-stroke impairment, some researchers have specifically studied SGER for cognitive impairment. Although SGER has been studied for ages, there is still no comprehensive review of exercise rehabilitation. For helping researchers better understand and develop SGERs, we analyzed the degree of the unhealthiness of different groups of people and proposed the classification of SGER. Then, based on this classification, we reviewed some studies of SGER in the past decade years.
\par In this paper, we illustrate the significance of the convergence of serious games and exercise rehabilitation. Then, the development of SGER based on hardware was reviewed. Next, we reviewed existing studies of SGER based on different groups of people and provided clear guidance for researchers. The rest of the paper is organized as follows: Section 2, we introduce the convergence of serious games and exercise rehabilitation. Next, Section 3 presents the serious games for different groups of people in exercise rehabilitation. Then, we proposed the related discussions and prospects in section 4. Finally, the paper ends with the conclusion section.

\section{The convergence of serious games and exercise rehabilitation}
\par Generally, people who used traditional exercise rehabilitation had difficulty maintaining interest in the rehabilitation process, so that rehabilitation was difficult to complete. And the serious game is a tool that is immersive and interactive and can stimulate people's imagination. Therefore, many researchers applied serious games to exercise rehabilitation and promoted the development of SGER. In this section, we introduced SGER and its development.
\subsection{The description of SGER}
Exercise rehabilitation is physical therapy with evidence-based sports science. It uses exercise to recover the patient. It mainly affects the whole body and part of the human body through nerve reflexes, neurohumoral factors, and biomechanical effects. Its clinical effects mainly include \cite{Handbook}: improving the body function, health status, and overall feeling of patients with injury, functional limitation or disability, preventing and minimizing potential damage, function limitation or disability, etc. In the clinical sub-field, it needs to set a baseline, such as physical condition, health information, previous exercise experience, etc. After the assessment, supervised rehabilitation training is used to achieve the set goals. The purpose of exercise rehabilitation is not only to restore muscles and joints, chronic pain or fatigue, neurological or metabolic diseases after surgery, but also to restore various psychological problems \cite{Xiao2017}. Besides, exercise rehabilitation can not only target patients but also sub-healthy people. It can improve or maintain sub-healthy people's body function, health, and mental state. Although traditional exercise rehabilitation is effective, its rehabilitation process has a single form and needs to be repeated. What’s more, the fatigue and pain in the traditional exercise rehabilitation process can also aggravate the player’s negative psychology. Luckily, using the characteristics of serious games may improve the player's psychology to help them complete rehabilitation training. Therefore, researchers tried to integrate serious games with exercise rehabilitation.
\par SGER is a type of serious game. It is not ordinary serious game but game that integrates serious game and exercises rehabilitation mechanisms. It makes use of the fun, interactivity, and immersion of the game to improve the single and repeated experience in the process of traditional exercise rehabilitation. It integrates the requirement of exercise rehabilitation into the game. Then it shows the exercises to players in the form of games to improve players' experience. This enables players to complete the exercise rehabilitation process better and ultimately improve the treatment effect. Niina Katajapuu et al. used SGER for exercise rehabilitation in the elderly in a controlled experiment and evaluated these elderly people with SPPB and Berg Balance Scale after the intervention, which proved that SGER is useful for improving the exercise function of the elderly \cite{8268221}. The review \cite{2017Active} by Mura, G et al. confirmed that SGER can improve the cognitive function of patients with neurological dysfunction. SGER can not only restore motor function and cognitive function alone but also restore them at the same time. A study by Alon Kalron et al. \cite{kalron2019virtual} showed that VR-based SGER is feasible and safe for multiple sclerosis patients with both dyskinesia and cognitive impairment. Therefore, SGER is an effective means of exercise rehabilitation.

\subsection{The development of SGER}
\begin{figure*}[h]
 \captionsetup{singlelinecheck=false}
 \centering
 \includegraphics[scale=0.5]{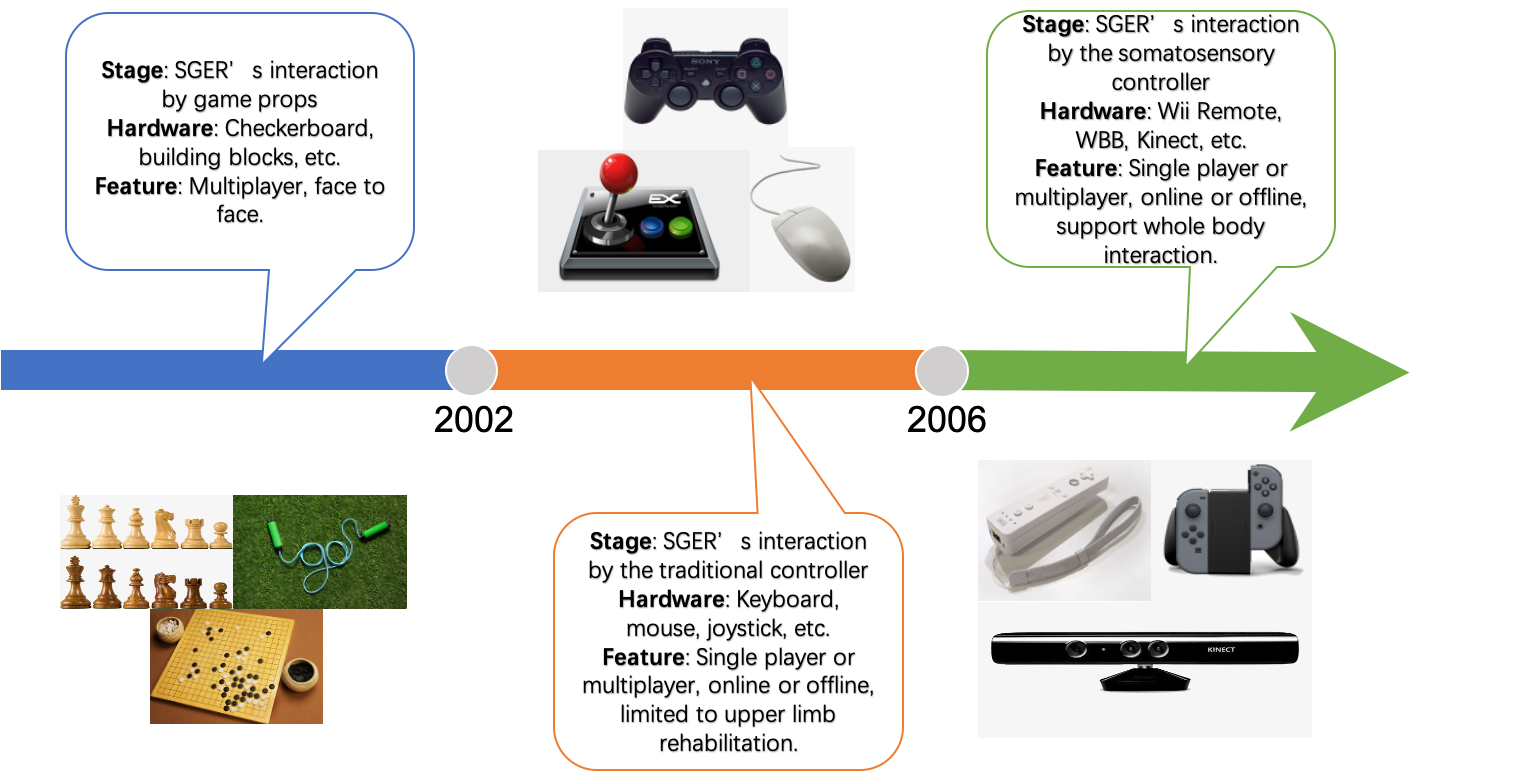}
 \caption{The development of SGER}
 \label{fig 1}
\end{figure*}

Understanding the development of SGER helps researchers have a clearer understanding of SGER. Therefore, as shown in Figure \ref{fig 1}, we reviewed the development of SGER from the perspective of game interaction method based on hardware. We divided SGER into three stages: SGER’s interaction by game props, SGER’s interaction by the traditional controller, and SGER’s interaction by the somatosensory controller. 

\subsubsection{The stage of SGER's interaction by the game props}
As early as the era without video games, there has been a combination of games and exercise rehabilitation. At this stage, players directly controlled the game props by hand. These games usually appeared in the form of a board, such as the Hi-Q mentioned in the research \cite{1993An} by Neistadet and others. This was a tabletop game. Players needed to remove as many stakes as possible by skipping adjacent stakes. It not only improved patients' cognitive function but also improved patients' upper limb motor function. Building block games could improve the fine motor skill of the human hands and the cognitive ability of thinking, imagination, and attention. Although the game at this time was not invented for exercise rehabilitation, it was enough to show that the convergence of exercise rehabilitation and early games was feasible and effective.
\par This stage’s games were multi-player games, which required players to face-to-face competition or collaboration or cooperation. Often, the games could not be played because there were not enough players, which made rehabilitation tasks impossible. Besides, the games had no clear purpose for exercise rehabilitation, so there were few cases where games were applied to exercise rehabilitation.
\subsubsection{The stage of SGER's interaction by the traditional controller}
With the development of computers, video games came into being. Video games are interactive games that run on the platform of electronic devices. They allowed a single player to play online or offline. Therefore, they overcame the drawback that the previous stage’s games can only be played by multiple players. What’s more, this stage's games displayed the game interface through the screen, and players manipulated the game through the traditional controller. This created a connection and a kind of feedback between players and electronic devices, that was, interactivity. Besides, the games could also simulate the real world or the world of thinking, which helped players strengthen their thinking and imagination. As shown in Table \ref{Table 1}, the control method of this stage's games was the traditional controller.
\begin{table*}[!t]
\centering
\caption{REPRESENTATIVE COMPANIES AND PRODUCTS OF TRADITIONAL CONTROLLERS}\label{Table 1}
\setlength{\tabcolsep}{3pt} 
\renewcommand{\arraystretch}{1.3} 
\begin{tabular}{|c|c|c|c|}
\hline
Company                    & Game console  & Controller          & Controller features                                                                                                          \\ \hline
Nintendo                   & Game Cube     & Wave Bird           & Wireless controller, Analog-stick                                                                                            \\ \hline
\multirow{3}{*}{Sony}      & PlayStation 2 & DualShock 2         & Vibration function, Analog-stick                                                                                             \\ \cline{2-4} 
                           & PlayStation 3 & DualShock 3         & Six-axis induction, Vibration function                                                                                       \\ \cline{2-4} 
                           & Playstation4  & DualShock 4         & \begin{tabular}[c]{@{}c@{}}Six-axis dynamic sensor, Vibration function, \\ Touch screen\end{tabular}                         \\ \hline
\multirow{3}{*}{Microsoft} & Xbox          & Duke                & Wired controller                                                                                                             \\ \cline{2-4} 
                           & Xbox 360      & Xbox 360 controller & Wireless controller, Vibration function                                                                                      \\ \cline{2-4} 
                           & Xbox One      & Xbox One controller & \begin{tabular}[c]{@{}c@{}}Vibration trigger, Vibration function, \\ Audio data interface, Bluetooth technology\end{tabular} \\ \hline
                           
\end{tabular}
\end{table*}
\par In the early days, both Nintendo’s Wave Bird controller and Sony’s Dual Shock 2 controller were equipped with analog sticks, which increased the player’s 3D experience during gaming. Sony's Dual Shock 2 controller was also equipped with 2 vibration motors, which gave players certain feedback in games. The feedback it provided could remind the patient of the wrong operation, which helped to the patient's cognitive rehabilitation. At present, Sony’s latest next-generation game console is PS4, and its main controller is DualShock 4, which supports dynamic sensing. It has an external touchpad and built-in six-axis dynamic sensors (three-axis gyroscope, three-axis accelerometer). Microsoft's latest next-generation game console is Xbox One X, and its controller has a touch feedback function by a vibration trigger and four Vibration motor.
\par Research on the impact of games on the brain jointly conducted by the Max Planck Institute for Human Development and St. Hedwig’s Hospital showed that some action games, like Super Mario, enabled the brain to control memory generation, spatial positioning, strategic planning, and subtle part of the field of motor function was expanded and activated \cite{SuperMario}, which helped patients recover. Music Conductor was a music game. Players needed to press buttons following the rhythm of the music in the game. Players needed to distinguish four different strengths for pressing the buttons. By playing this game, players could train control of hand power and cognitive ability. Reinkensmeyer et al. \cite{1031978} modified the classic arcade game "Breakout!" to a therapy game "Breakout Therapy". In this game, the player controls the joystick to hit the game target to complete the rehabilitation task. Players participating in the experiment said that by playing the game, their arm movement ability has been improved.
\par Although this stage’s games overcame the limitation of the number of people, there were still many limitations. Such as, the game operation mode was limited to the traditional controller, players must stare at the screen for a long time when playing games which is harmful to eyesight, the movement range is small lead to players cannot train their overall movement ability and balance ability when playing games, etc.
\begin{table*}[!t]
\centering
\caption{REPRESENTATIVE COMPANIES AND PRODUCTS OF SOMATOSENSORY CONTROLLERS}
\label{Table 2}
\setlength{\tabcolsep}{3pt} 
\renewcommand{\arraystretch}{1.3} 
\begin{tabular}{|c|c|c|c|}
\hline
Company                   & Game console         & Controller        & Controller   features                                                                                                                                  \\ \hline
\multirow{3}{*}{Nintendo} & \multirow{2}{*}{Wii} & Wii Remote        & \begin{tabular}[c]{@{}c@{}}Pointing and positioning,  \\ Motion-sensing (detection of controller stereo tilt changes)\end{tabular}                     \\ \cline{3-4} 
                          &                      & Wii Balance Board & Pressure measurement, The center of gravity measurement                                                                                                \\ \cline{2-4} 
                          & Switch               & Joy-Con           & \begin{tabular}[c]{@{}c@{}}Acceleration sensor, Gyroscope, Infrared depth sensor, \\ Near field communication, High-precision vibration\end{tabular}   \\ \hline
\multirow{2}{*}{Sony}     & PlayStation2         & Eye Toy           & Camera                                                                                                                                                 \\ \cline{2-4} 
                          & PlayStation3         & PS Move           & \begin{tabular}[c]{@{}c@{}}Three-axis gyroscope, Three-axis accelerometer,  \\ Earth magnetic field sensor, Vibration feedback, Bluetooth\end{tabular} \\ \hline
Microsoft                 & Xbox360              & Kinect            & \begin{tabular}[c]{@{}c@{}}3D-camera, Motion capture, Image recognition, \\ Microphone input, Voice recognition, Social interaction\end{tabular}       \\ \hline
HTC                 & -----              & HTC Vive Tracker            & \begin{tabular}[c]{@{}c@{}}6-DOF tracking, Motion capture\end{tabular}       \\ \hline
Thalmic Labs                 & -----              & MYO Wristband            & \begin{tabular}[c]{@{}c@{}}Fast gesture capture,\\ Capture the bio-electricity of arm muscles\end{tabular}       \\ \hline
Leap                 & -----              & Leap Motion            & \begin{tabular}[c]{@{}c@{}}3D-camera, Motion-sensing, \\8 cubic feet of interactive range 3D scape\end{tabular}       \\ \hline

\end{tabular}
\end{table*}

\subsubsection{The stage of SGER's interaction by the somatosensory controller}
With the further upgrade of game controllers and the need to enhance the player experience, active video games appear in people's field of vision. Active video game (also called Exergame) was a kind of video game that was experienced with the body. The biggest difference between it and last stage's SGER was the game’s control method. This stage’s games were video games that players used body movement to operate. Obviously, this required players to exercise. It just so happened that the realization of exercise rehabilitation was also exercise, so it was more suitable for combining with exercise rehabilitation than previous games. As we expected, more and more therapists recommended that people who need exercise rehabilitation to use SGER for rehabilitation training. This stage’s games not only broke through the limitations of the first stage’s games, but also broke through the limitations of the second stage’s games that have limited motion, insufficient player experience, and inconvenient control equipment. Table \ref{Table 2} shows some of this stage’s controllers.
\par Wii is the seventh generation game console developed by Nintendo. The controller shipped with it is called the Wii Remote, which is a stick-shaped hand-held portable pointing device. It can control the cursor on the screen like a mouse. It can detect movement in three-dimensional space. The combination of the two can achieve the so-called "somatosensory operation" \cite{WiiRemote}. Later, Nintendo introduced a new controller Wii Balance Board (WBB), which has functions of pressure measurement and center of gravity measurement. Many researchers apply it to balance rehabilitation. PS is a home game console sold by Sony Corporation of Japan. PlayStation Move (PS Move) is a new type of controller for PS3, which has six-axis sensing, vibration, and somatosensory functions. PS Move not only recognizes the movements up and down, left and right, but also senses changes in the angle of the wrist. Besides, PS Move needs to be used with PlayStation Eye. Xbox 360 is a home game console developed by Microsoft. Kinect is a 3D motion camera that completely subverts the single operation of the game. Kinect has functions such as dynamic capture, influence identification, microphone input, voice recognition, and social interaction. It can directly use the 3D-camera to capture the player's every movement, rather than limited to hands. Many researchers used Kinect to track the patient's bone movement information to monitor the rehabilitation process in real-time and specify a better rehabilitation plan. HTC Vive Tracker is part of the HTC Vive, which can track motion with six degrees of freedom. It is often used by researchers to capture the axial rotation of the player's forearm or leg. Leap Motion is a somatosensory controller for gesture tracking. It has 8 cubic feet of interactive range 3D scape and can fully track the movement of the arm and palm. Therefore, researchers often use it in the process of upper limb fine motor rehabilitation. The MYO wristband is also a somatosensory controller used to track gestures. Compared with Kinect and Leap Motion, it is not limited by specific venues and has a faster speed of gesture capture, but the accuracy and richness of gestures are less than Leap Motion. It is a good choice if an experiment need to collect electrical signals from muscles.
\par In this stage, there are many SGERs. Such as CatEye GameBike \cite{1707518} converts the exercise bike into a video game controller when it is connected to the PlayStation2. Then, the player can exercise while playing various games (racing, speedboat, motorcycle racing, etc.) on the computer. Cube Destroyer \cite{8009394} is also a SGER controlled by body movement. In this game, the player uses his/her knee extension to operate the game avatar. Not only can play game immersively but also perform exercise rehabilitation.
\par With the advancement of game controllers, the related technology of the game also advances. Such as virtual reality, augmented reality, robot technology, artificial intelligence (AI), telerehabilitation, and so on. Advances in these technologies have promoted the development of SGER. Such as Moldovan et al. \cite{7995427} used the MIRA platform game, which is equipped with virtual reality, to treat a patient’s dyskinesia. After treatment, the patient expressed affirmation of the SGER and wanted to continue playing the SGER; Pirovano et al. \cite{6967825} used AI to realize a virtual therapist, who can correctly judge the patient’s condition.
\par Even so, this stage’s games still have many limitations. Since the games were controlled by body movement, players needed to have a certain degree of exercise ability. However, some patients couldn't meet this standard. Therefore, researchers should pay attention to the use of robotic exoskeleton technology. What’s more, the interactive mode of these games also determined the high immersion of the game. Players may forget the surrounding environment during exercise, which can lead to falls and other accidents. Therefore, the choice of venue should also be paid attention to.

\section{Serious games for different groups of people in exercise rehabilitation}
In this section, a classification based on different groups of people is given for SGER and some relevant studies in the last decade are reviewed based on this classification. According to our survey, the group of exercise rehabilitation is divided into the sub-healthy group and patient group. Sub-healthy people are in chaos physically or mentally, but they have no obvious pathological features \cite{Subhealth1}. For them, the use of SGER can restore the physical and mental state to prevent the occurrence of pathological changes. But, the patient's body or psychology has pathological characteristics. In this review, they are patients with dyskinesia or cognitive impairment or both. For them, the use of SGER can help them treat the condition. Therefore, we divide SGERs into SGER for sub-healthy people and SGER for patients according to the degrees of the unhealthiness of different groups. Considering that current researches' focus are on patients, we further divide SGER for patients into SGER for motor rehabilitation, SGER for cognitive rehabilitation, and SGER for integrated rehabilitation. Finally, based on this classification, some relevant studies in the last decade are reviewed to help researchers develop better SGERs.
\subsection{Serious games for sub-healthy people in exercise rehabilitation}
In today's society, many sub-healthy people felt confused and frustrated due to the decline of their physical fitness or cognitive level or poor mental state. Failure to pay attention to these problems may lead to pathological changes in their body and mind. Such as, the long-term poor mental state leads to depression, long-term lack of exercise leads to obesity, loss of exercise capacity, or various other complications. These will have a huge impact on their family. Therefore, more and more sub-healthy people choose to use SGER to restore their physical and mental states to prevent diseases.
\par Staiano et al. \cite{2013Adolescent} studied whether exergames can reduce weight and improve the mental state of overweight or obese adolescents. The growth curve model after the intervention showed that exergames (especially the cooperative exergames) can reduce the weight of overweight or obese adolescents and increase their self-esteem, self-efficacy, and peer support. The review by Boulos \cite{2012Xbox} showed that exergames based on Kinect had a positive effect on the physical and mental states of all age groups. SGERs based on the family had great limitations in exercise scope and intensity. Mobile-based SGERs lacked immersion and interactivity of family-based SGERs, while gamified traditional exercise rehabilitation lacked interest. Asynchronous games \cite{8882471} may be a good choice. For example, Solitaire Fitness was a card game designed to increase the fitness interest of the elderly and enhance the cognitive and athletic abilities of the elderly. Different from other card games, players could gain more game skills through the exercise to help them win the game.
\par Although there were many SGERs (Wii fit, Just dance, Ring Fit Adventure, etc.) on the market, there were few related results of research about sub-healthy people using SGERs. The reasons for this situation may be: 1) The extent of their need for exercise rehabilitation games is much lower than that of patients, which has not attracted widespread attention from researchers; 2) Some sub-healthy people who need to use SGER are very concerned about their health. They will actively try to play SGER to avoid adverse consequences, so researchers pay less attention to them; 3) Some sub-healthy people who need SGER exercise neglect their own health and fail to use SGER to recover in time so that their body occur pathological changes. Therefore, this part of the population will appear in group of patients.
\subsection{Serious games for patients in exercise rehabilitation}
\begin{figure*}[h]
 \captionsetup{singlelinecheck=false}
 \centering
 \includegraphics[scale=0.5]{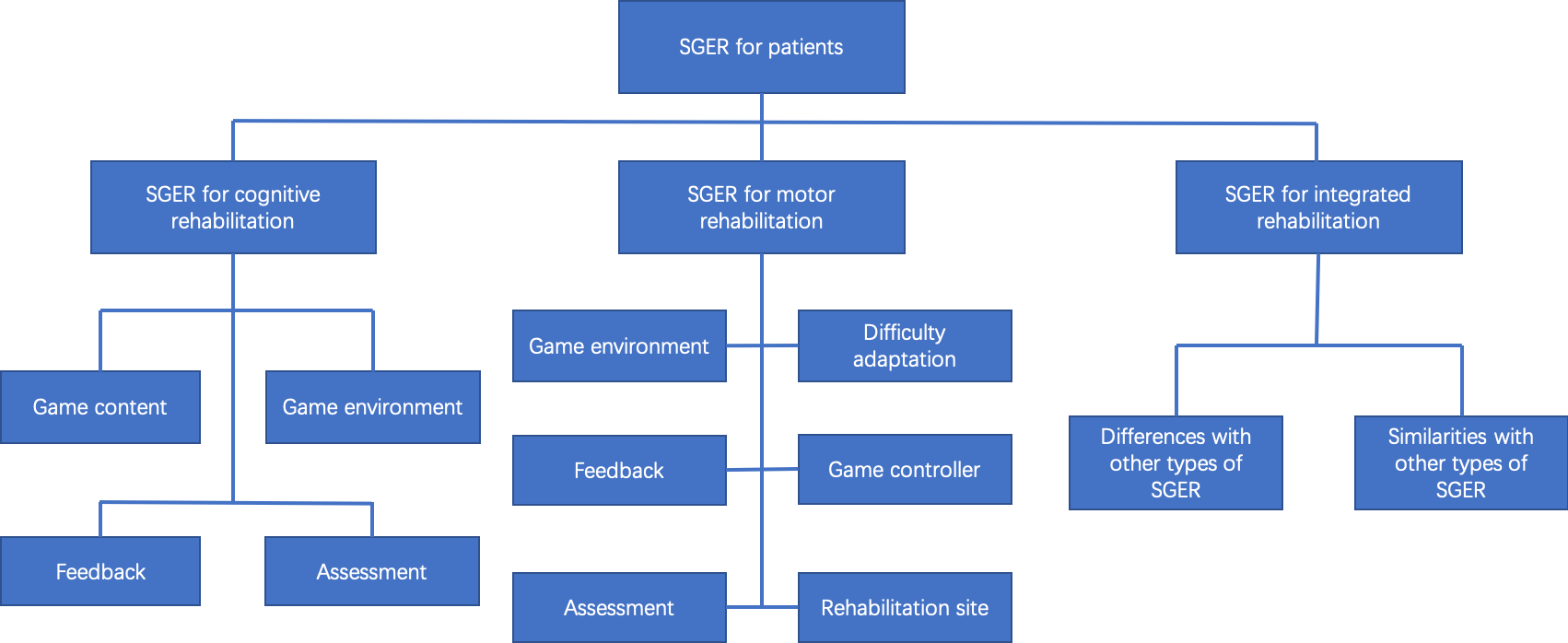}
 \caption{Knowledge framework of SGER for patients}
 \label{fig 2}
\end{figure*}

Patients who need exercise rehabilitation often have varying degrees of cognitive impairment and dyskinesia. The cases studied in this review hardly include patients with severe cognitive impairment or severe dyskinesia. Because patients with severe cognitive impairment often cannot understand SGERs, patients with severe dyskinesia do not have the athletic ability to complete exercise rehabilitation. According to the different combinations of the types of rehabilitation abilities that patients need, we divide them into three categories: SGER for cognitive rehabilitation, SGER for motor rehabilitation, and SGER for integrated rehabilitation. Besides, we also summarized the knowledge framework of this part so that researchers can better understand the paper, as shown in Figure \ref{fig 2}.

~\\
\subsubsection{SGER for cognitive rehabilitation}
Cognition is the intelligent process of human body recognition and knowledge acquisition. It includes memory, language, visual space, execution, calculation and comprehension judgment. Cognitive impairment refers to the impairment of one or more of the above-mentioned cognitive functions and affects the individual's daily or social abilities. The clinical manifestations of it are perception barriers, memory disorders, and thinking disorders. Diseases associated with cognitive impairment usually include dyslexia, stroke, multiple sclerosis, depression, and so on. Regarding cognitive impairment, many researchers have conducted researches on SGER for cognitive rehabilitation. Below we review these studies.  
~\\
\par \textbf{Game content}
\par Different game content has different effects on cognitive rehabilitation. Hung, JW et al. \cite{2017Cognitive} compared the effects of Wii Fit, Tetrax biofeedback system games, and conventional weightlifting balance training on improving and maintaining the cognitive function of patients. Among them, Wii Fit was a commercial electronic game, Tetrax biofeedback game was a game that focused on balance rehabilitation, and conventional weightlifting balance training was traditional exercise rehabilitation. The result of the research showed the gains of the Wii Fit group in the abstract/judgment domain and language domain were higher than the other two groups. The author pointed out that this may be attributed to the 5 Wii fit games selected 1) had better visual and auditory experience, which helped improve the patient’s attention, thereby improving reading ability; 2) using planning and decision-making cognitive process and keeping patients in a state of higher cognitive load helps to enhance abstract thinking and judgment. Also, Catalan et al. \cite{2016Phantom} used a racing game with a phantom limb movement to perform rehabilitation experiments on patients with phantom limb pain (with perception impairment) and achieved positive results. Besides, Franceschini et al. \cite{franceschini2013action} compared the efficacy of action video games (AVG) and nonaction video games (NAVG) and found that AVG can improve the reading ability of players. Further research found that AVG improves the ability of reading by increasing the player's spatial and temporal attention. Like the three examples mentioned earlier, the first one is to add a better visual and auditory experience to the game, which has the effect of improving the patient’s attention; the first one also adds the planning and decision-making process to the game to improve the thinking logic and judgment ability; the second is to add phantom limb movement to the game, playing a role in the treatment of phantom limb pain; the third one used AVG to improve the patient’s reading ability. Therefore, researchers should explore the effects of different game content on cognitive rehabilitation to provide more disease-specific targeting for SGER’s development.
~\\ 
\par \textbf{Game environment}
\par VR and AR environments are suitable for players to perform SGER for cognitive rehabilitation. Active Brain Trainer (ABT) \cite{8007530} was a platform based on VR. Shochat et al. used the platform’s games to treat patients with acquired brain injury. After treatment, the patient expressed satisfaction with ABT and hoped to continue using it in subsequent treatments. R. Ocampo and M. Tavakoli compared the user performance of AR and non-immersive VR in \cite{8604049}. In the experiment, patients were required to play three SGERs (Snapping, Catching, and Ball Dropping). These three games tested the patients' cognitive abilities such as spatial perception, attention, and accuracy. Patients' game scores in the AR environment were usually higher than the scores in the VR environment, and there was no significant difference between subjects with the same group of cognitive load. Therefore, when developing SGERs, researchers should try to choose VR or AR as the game environment.
~\\
\par \textbf{Feedback}
\par Feedback is an important part of SGER for cognitive rehabilitation. There are visual feedback, haptic feedback, and auditory feedback. Different feedback contributes to the rehabilitation of different cognitive functions. Saegusa et al. \cite{6974060} compared the effects of three cognitive assistance methods on subjects in exercise tasks. The results showed that auditory feedback was more effective than visual feedback for grasping the opportunity of performing the exercise. For spatial recognition and perception, visual feedback was more effective than auditory feedback. Audio-visual feedback was usually more effective than auditory feedback and visual feedback. Ma, WQ et al. \cite{ma2017comparison} compared the effects of visual feedback, haptic feedback, and visual-haptic feedback. The results showed that visual feedback helped to decision-making. Haptic feedback provided more cognitive skills' training than visual feedback. However, visual-haptic feedback didn't provide as much help as the previous two. This may be because the visual-haptic feedback brought more cognitive load or distraction to the subjects. Therefore, when researchers develop SGER for cognitive rehabilitation, they should not only pay attention to the choice of feedback form but also pay attention to the degree of feedback.
~\\  
~\\
\par \textbf{Assessment}
\par Assessment is the key for researchers to measure the patient's effect of rehabilitation. The most common assessment method used by researchers is the existing clinical scale. Before the execution of SGER, \cite{8007530} used the Montreal Cognitive Assessment(MoCA) to assess the patient’s level of cognitive impairment. After the execution of SGER, \cite{2017Cognitive} used Cognitive Abilities Screening Instrument Chinese version (CASI C-2.0) and Berg Balance Scale (BBS) to assess the status of patients’ cognitive rehabilitation; \cite{2019Improving} used subtests of the Illinois Test for Psycholinguistic Aptitudes (ITPA) to evaluate the patient’s visual comprehension, visual association, visual integration, and visual-motor sequential memory; \cite{8007530} used Executive Function Performance Test (EFPT), Executive Abilities: Measures and Instruments for Neurobehavioral Evaluation and Research (NIH-EXAMINER), and Dysexecutive Questionnaire (DEX) to assess the patient’s executive functions. And, sensor-based assessment methods are also effective. In the study by Ciman et al. \cite{ciman2018serious}, an eye tracker was used to track the eye movement of patients while playing games to collect data on eye gaze and to further evaluate the patient's recovery status. The experiment of Phillipp Anders et al. \cite{anders2018exergames} recorded the player's EEG. By observing the absolute theta power in the frontal cluster in the EEG, the changes in the patient's cognitive needs could be learned. Besides, AI and computational intelligence (CI) can be used to assess patients, too. Jung et al. used AI technology to propose a model called Neuro-World \cite{8834484} for predicting the patient’s rehabilitation effects. The model training model used the CFS algorithm. There was a 9.8\% error between the final predicted result and the actual result. Therefore, we recommend that researchers explore more feasible evaluation methods (for example, using deep learning models for patient evaluation) on the basis of ensuring accurate patient assessments, so that the existing assessment methods are more effective and comprehensive.
~\\
\subsubsection{SGER for motor rehabilitation}
There are two types of motor functions: voluntary movement and involuntary movement. Voluntary movement is conscious and can be carried out with one's own will, also known as an autonomous movement. Involuntary movement refers to the movement of the heart muscle and smooth muscle controlled by the visceral motor nerve and the vascular motor nerve. It is a movement that is unconscious and not controlled by your own will. Dyskinesia mainly refers to the disorder of autonomous movement ability, the movement is incoherent, unable to complete, or completely unable to exercise at will. The main clinical manifestations are involuntary exercise, lack of exercise or slowness without paralysis, abnormal posture, and muscle tone. Diseases associated with dyskinesia usually include stroke, cerebral palsy, Parkinson's disease, amputation, etc. With regard to dyskinesia, many researchers have conducted researches on SGER for motor rehabilitation. Below we review these studies.
~\\ 
\par \textbf{Game environment}
\par A few years ago, virtual reality has emerged in the field of SGER for motor rehabilitation. In the research of Moldovan et al. \cite{7995427}, the MIRA platform games based on virtual reality were selected to train the gross motor function of the upper limb and the balance and the coordination of body. An 81-year-old post-stroke patient successfully improved his upper limb motor ability, balance, and coordination after treatment. The patient said that the rehabilitation under the virtual environment did not produce adverse reactions and liked to play MIRA interactive video games. Motion Rehab AVE 3D \cite{trombetta2017motion} compared non-immersive VR and immersive VR. Experiments showed that Motion Rehab AVE 3D was more suitable for non-immersive VR because it was a third-person game. Also, Trombetta pointed out that immersive VR was more suitable for the first-person SGER. Mazhar Eisapour et al. \cite{eisapour2020participatory} compared immersive VR rehabilitation games with traditional exercise rehabilitation, and the results showed that immersive VR games could help people with dementia participate in exercise rehabilitation. E.D.Oña et al. designed the Box and Blocks Test \cite{8882472} based on VR and proved that the use of virtual reality systems could effectively and reliably assessed the patient's hand movement ability. Compared with the real environment, the VR environment was a more friendly environment of assessment. As the advanced stage of VR, AR enables players to directly interact with the fusion of the physical world and the virtual world to bring players a more realistic experience. ChiroChroma \cite{8088461} was an AR-based SGER, the game was divided into two stages. In the first stage, the game provided the player with a 2D drawing. The player performed a pinching motion to select the colored area and color it. This stage could exercise the patient's hand movement ability. In the second stage, the virtual target was positioned at a fixed position in front of the virtual camera. The player needed to drag an object to the HMD device and released it to hit the target. The score depended on the hit area. This stage could exercise the patient's arm movement ability. The questionnaire survey revealed that although the AR system’s usability score was below average, players still said they felt immersed in it and were driven to complete the game.
~\\ 
\par \textbf{Difficulty adaptation}
\par In the actual exercise rehabilitation process, the therapist will change the difficulty of the action of rehabilitation according to the patient's real-time performance. Exercise rehabilitation in SGER should be the same. To achieve difficulty adaptation, Nirme et al. \cite{6091665} applied and compared a random line search algorithm and a predictive search algorithm in the spheroids game. The results showed that both algorithms have effectively completed the difficulty adaptation and are robust. It was worth mentioning that the random line search algorithm will increase the error as the game difficulty parameter increases, and this situation will not occur in the prediction search algorithm. Zainal et al. \cite{9073659} proposed a new scoring mechanism using K-means clustering, with the purpose of dynamically changing the difficulty of the game for patients. They successfully used the algorithm to complete the score calculation and difficulty recommendation. Besides, Esfahlani et al. \cite{8001435} 1) used the IK algorithm and dynamics method to simulate the movement of the patient’s upper limbs; 2) developed a multi-input single-output fuzzy system, which could imitate human thinking, and then made decisions based on the player's performance to provide patients with personalized autonomous rehabilitation programs (included difficulty adaptation). The above three examples all applied CI and AI to difficulty adaptation and have achieved positive results. Therefore, CI and AI are powerful tools to achieve difficulty adaptation to ensure the efficacy and safety of rehabilitation.
~\\ 
\par \textbf{Feedback}
\par Feedback is an important part of SGER for motor rehabilitation. It can not only guide players to the correct exercise but also increase the simulation of the game. Kommalapati and Michmizos \cite{7592058} speculated that observing one's own activity activates the "mirror neuron system", which help to motor rehabilitation. Based on this, they proposed a visual feedback for children with severe dyskinesia called “Visual Gain”. This feedback gave the player an illusion that he/she could better complete the rehabilitation task, so that the player's motion response quality and participation were improved, thereby improving the rehabilitation effect. IGER \cite{6967825} used two types of feedback, visual feedback, and auditory feedback. Its visual feedback was to mark the body parts of the game avatar with colors. When there was a problem with the player's actions, the body parts of the game avatar would appear in conspicuous colors to warn the player. This allowed the player to pay attention to and correct their wrong actions in time to improve the healing effect. And, it provided auditory feedback in three ways, namely a humanoid virtual therapist, piggy, and a real therapist. The same was that the purpose of the three was to guide the player to perform rehabilitation exercises. The difference was that the author emphasized the fidelity of the humanoid virtual therapist and the fun of piglet. Besides, Khor et al. \cite{6775492} developed a haptic interface based on the Compact Rehabilitation Robot (CR2). The haptic interface could provide players with haptic feedback and visual feedback under the control of the control algorithm. When the player hit a wall in the game, the tactile feedback would give the player a force feedback, making the player feel a real sense of collision. When the player received raindrops in the game, the haptic feedback would increase the resistance as the quality of the water in the cup in the game increases. Using this method can increase the simulation of the game and make the player feel immersive.
~\\ 
\par \textbf{Game controller}
\par SGER for motor rehabilitation has a wide selection of controllers, and the correct choice is the basis for successful rehabilitation. Esfahlani et al. \cite{8001435} used the Kinect bone tracking system, which can detect the position and direction of a single joint and the speed of joint motion. By using Kinect, Esfahlani and others successfully extracted the patient's movement information to evaluate the treatment effect. R. Alexandre and O. Postolache developed a wearable smart glove in \cite{8538058}. The smart glove was equipped with a multi-functional sensor and used an Arduino Nano microcontroller to perform data collection. After initial processing, the collected data was transmitted to the game client through Bluetooth. With the help of the Internet of Things architecture, the therapist finally realized the remote assessment of patients to ensure that the rehabilitation plan adapts to the needs of the patients. In the experiment of Elnaggar and Reichardt \cite{7881304}, considering that patients may not be able to wear wearable devices due to lack of fingers or pain, the lightness and portability of game hardware, and the high accuracy of data acquisition requirements, the Leap Motion device was finally selected. They adapted an Angry bird game. Players needed to hold the handle firmly and move their hands up, down, left, and right to change the shooting direction of the birds. The results showed that the angle of the finger joints could be increased by playing the Angry bird game. Cube Destroyer \cite{8009394} was a game based on the concept of infinite runners. The player needed to control the spacecraft moving forward at a constant speed through the extension and contraction of the knee extensor muscles to move up and down, so as to traverse rings and cubes at different heights and avoid obstacles. Bulea et al. provided patients with a robotic exoskeleton in the experiment to assist patients' knee extension. The patients' EEG results showed that the use of robotic exoskeleton-assisted exercise would not simply transfer the exercise that patients should have done to the exoskeleton, and patients were still engaged in the exercise they should do.
~\\ 
\par \textbf{Assessment}
\par If there are no appropriate assessment methods, then the patient's rehabilitation effect will not be accurately assessed, so that the patient participates in invalid or even counterproductive rehabilitation exercises. Firstly, the existing clinical assessment scales are an effective way to assess patients, and many researchers use them to evaluate patients. Such as \cite{8994565}, \cite{7995427}, \cite{9069313} both used Fugl-Meyer assessment scale to assess patients' motor ability. And, the MAL scale was used in \cite{8994565}, the Berg Balance Scale and ARAT were used in \cite{7995427}. Secondly, sensor-based assessment methods have the potential to complement existing assessment techniques \cite{6289368}. For example, Synnott et al. \cite{6289368} proposed and used a sensor-based assessment method called WiiPD. Experimental results showed that this remote assessment method was an effective method to assess the condition of PD patients. Thirdly, the game score can also be used as an indicator to assess recovery. In a study \cite{9069313}, Noveletto et al. proposed a game scoring system for evaluating muscle control. The correlation of the QFG Score, HSG Score, and FMA-LE showed that game indicators accurately correspond to clinical results. Besides, the optimization of the assessment method is also important. Rodriguez-de-Pablo et al. \cite{7319424} explored the correlation between the different components of the ArmAssist assessment (AAA) and certain current clinical scales. The study found that ROM, COM, ROF game scores in AAA have a moderately significant correlation with all clinical scores. The AAA total score component was set to a combination of ROM, COM, and ROF, which optimizes the assessment method of patients' sports scores, and contributes to remote assessment and automatic adjustment therapy.
~\\ 
\par \textbf{Rehabilitation site}
\par With the development of the Internet and the Internet of Things, telerehabilitation for SGER for motor rehabilitation has gradually become a very popular technology. MyoBeatz \cite{8882432} was a telerehabilitation game based on music. In the game, players need to contract their muscles according to the rhythm of the song and grab the arrow that rolls up from below. Different muscle contraction strength and duration correspond to different functions. Its game system components could be taken home and presented in a compact form for easy use. The user Mobile Application Rating Scale (uMARS) results showed that the game has a high impact on the target population for home training of EMG. The results of clinical parameters showed that the muscular endurance of all patients was significantly improved, and the neuromuscular control ability around the amputation site was also significantly improved. These showed that the game was very effective for home training of prosthetic control for the target population. A patient three years after a stroke using the combination of Jintronix©.exergame and Reacts© applications \cite{8994565} for telerehabilitation training for 2 months. In the fourth week, the patient developed an abnormality of increased upper limb heaviness. The clinician guided the patient remotely and relieved the patient's discomfort. Besides, there was no bad situation, so it was safe to play SGER remotely. The results of the Fugl-Meyer and MAL assessment scales showed that training combined with the remote rehabilitation system and SGER can improve the motor function after stroke, which showed that playing SGER remotely was effective. Pirovano et al. \cite{6967825} applied CI to telerehabilitation, which further improved the safety and effectiveness of telerehabilitation. IGER was a rehabilitation game engine that uses CI to dynamically control the difficulty of the game while monitoring the player's behavior and providing he/she with effective feedback. Among them, the dynamic difficulty adaptation technology used a Bayesian framework, and the feedback used a novel color coding. Also, its fuzzy system allowed the therapist's knowledge to be encoded intuitively to ensure that the virtual therapist can make correct judgments about the patient's condition.
~\\
\subsubsection{SGER for integrated rehabilitation}
Some studies have shown that many patients have both cognitive impairment and dyskinesia. They need an integrated rehabilitation program to treat themselves. Common diseases include stroke, aging, Parkinson's disease, autism, etc. With regard to this type's impairment, many researchers have conducted researches on SGER for integrated rehabilitation. When reviewing these literatures, we found that SGER for integrated rehabilitation and SGER for cognitive (motor) rehabilitation have both similarities and differences. Below we will review these literatures.
~\\ 
\par \textbf{Differences}
\par The first difference is the severity of the illness. We found that SGER for integrated rehabilitation targets patients with more complex conditions, and they were often in the middle and late stages of the condition, which made the treatment more difficult and made it easier for patients to give up treatment. Therefore, SGER for integrated rehabilitation should better meet the psychological needs of patients. Fruites \cite{7800105} was a Kinect-based SGER for integrated rehabilitation tailored for children with autism. The style of the game was cartoon style, which may help to increase the interest of children with autism. The game contained five modes: "Sorting", "Mathematic", "Catching", "Imitation" and "Seeking". These five game modes all included cognitive rehabilitation and motor rehabilitation contents. For example, "Sorting", the player needed to use gestures to control the palm of the game to pick the fruit from the tree and put the fruit into the designated basket. In high difficulty, players also needed to avoid distractions. As the number of games played increased, the attention of children with autism who participated in the test decreased. Therefore, when developing SGER for integrated rehabilitation for children with autism, researchers should pay attention to playtime and control of the fun. The balloon goon game \cite{7551908} was also a cartoon-style SGER for integrated rehabilitation based on Kinect. It was used for the integrated rehabilitation of patients with Parkinson's disease. Because the conventional game menu bar required patients to accurately make the correct choice from many options, which may reduce the patient’s gaming experience and aggravate cognitive impairments. This game specifically set two exercises (called Navigation Gestures) for PD patients to control the game menu. This enlightens us: 1) the integration of convenience mechanisms for patients and rehabilitation may be a good method; 2) rehabilitation tasks can appear anywhere in the game, rather than limited to the process of playing games.
\par The second difference is the game’s type of rehabilitation tasks. The rehabilitation tasks of SGER for cognitive/motor rehabilitation is cognitive/motor function rehabilitation tasks. But, the rehabilitation tasks of SGER for integrated rehabilitation is the dual cognitive task \cite{8405454}. The dual cognitive task is designed to treat cognitive impairment and dyskinesia at the same time. The study by Kannan et al. \cite{kannan2019cognitive} proved the effectiveness of dual cognitive tasks for comprehensive rehabilitation. Pichierri et al. \cite{pichierri2012cognitive} used the Dance video game with dual cognitive tasks to exercise the elderly in the dance group. The results showed that the brisk walking performance of the dance group that received additional cognitive task training was significantly improved compared with the control group that only performed resistance training and balance training. Besides, HaShim et al. \cite{8405454} proposed a virtual reality game framework for the dual cognitive tasks of stroke rehabilitation. The framework required SGER to choose adventure as the game type, used the Prisoner Dilemma game theory, used the Interactive Cognitive-Motor Training technique, and used the virtual reality environment. This framework could effectively increase the interest of patients. However, as far as we know, there were few SGERs for dual cognitive tasks. Therefore, researchers should develop more SGERs with dual cognitive tasks to support patients' integrated rehabilitation.
~\\
\par \textbf{Similarities}
\par Similar to SGER for cognitive/motor rehabilitation, SGER for integrated rehabilitation also widely applies advanced technologies such as VR, AR, and difficulty adaptation. ROBiGAME \cite{7939262} was a SGER for integrated rehabilitation used on the robot end-effector REAplan. REAplan could adjust the control difficulty of the handle in real-time according to the patient's performance in the game to ensure the safety and effectiveness of the patient's upper limb exercise rehabilitation process. If the patient didn't perform well in the game, REAplan would secretly provide the patient with different degrees of exercise assistance to help the patient complete the rehabilitation training and improve the patient's self-confidence. Also, when the patient felt the game was difficult, it could provide prompts to the patient through its display to help the patient's cognitive rehabilitation. In the experiment of Bortone et al. \cite{8384290}, Moneybox Game trained grasping, hand extension, and the combination of internal rotation and supination of the forearm, and Labyrinth Game trained arm movement and coordination of eyes and hands. In this process, new wearable devices were used to monitor the movement of patients' fingers and HMD was used to provide patients with an immersive VR experience. Not only difficulty adaptation technology and VR technology were applied to SGER for integrated rehabilitation, but also AR technology had examples of being used. For example, Add VS Sub and Stroop Game \cite{8771093} were SGERs based on AR aimed at integrated rehabilitation. After using them, patients improved their athletic ability and self-confidence. Besides, telerehabilitation technology is also a cutting-edge game technology. Ghisio et al. \cite{7325501} developed three comprehensive telerehabilitation SGERs based on an open telerehabilitation platform, namely CULT, HTCT, and CERT. Based on the two workstations (hospital workstation and home workstation) of the rehabilitation platform, the telerehabilitation program of SGER for integrated rehabilitation was realized. 

\section{Prospect and Discussion}
The application and development of serious games in the field of exercise rehabilitation have important significance and broad development space. A lot of researches by researchers has promoted the integration of serious games and exercise rehabilitation. Whether it is the game where players interact with game props, the game that players interact by the traditional controller, or the game that players interact  by the somatosensory controller, it can be applied to exercise rehabilitation. Besides, many new technologies are used to develop SGERs, such as robotics, VR/AR, AI and CI, and so on. However, SGERs are still facing challenges in game development, technological innovation, and social popularization. Taking these into consideration, we first discussed SGER from SGER’ technology development (as shown in Table \ref{Table 3}) perspective, and then discussed SGER from the perspective of functional design and social popularization:
\begin{table*}[!t]
\setlength{\tabcolsep}{3pt} 
\renewcommand{\arraystretch}{1.3} 
\caption{TECHNOLOGIES OF SERIOUS GAMES FOR EXERCISE REHABILITATION}
\label{Table 3}
\centering
\begin{tabular}{|c|c|c|}
\hline
SGER technology       & Application                                                                                                                                                                                        & literatures                                                                                                                                                                       \\ \hline
VR                    & \begin{tabular}[c]{@{}c@{}}Virtual scene, Improving game simulation, \\ Increasing immersion and interaction\end{tabular}                                                                          & \begin{tabular}[c]{@{}c@{}}\cite{7995427}\cite{8007530}\cite{8604049}\cite{trombetta2017motion}\\ \cite{eisapour2020participatory}\cite{8882472}\cite{8384290}\end{tabular}                                                                               \\ \hline
AR                    & \begin{tabular}[c]{@{}c@{}}Combines real scenes and virtual scenes, \\ Improving game simulation, \\ Increasing immersion and interaction\end{tabular}                                             & \cite{8604049}\cite{8088461}\cite{8771093}                                                                                                                                                          \\ \hline
Difficulty adaptation & \begin{tabular}[c]{@{}c@{}}Adjusting the difficulty of the game, \\ Ensuring the safety and effectiveness of rehabilitation, \\ Improving the game experience of players\end{tabular}              & \cite{6091665}\cite{9073659}\cite{8001435}\cite{7939262}                                                                                                                                                  \\ \hline
Feedback              & \begin{tabular}[c]{@{}c@{}}Assisting exercises, Tips, \\ Warnings, Improving game simulation, \\ Increasing immersion and interaction\end{tabular}                                                 & \begin{tabular}[c]{@{}c@{}}\cite{6967825}\cite{6974060}\cite{ma2017comparison}\cite{7592058}\\ \cite{6775492}\end{tabular}                                                                                               \\ \hline
Assessment            & \begin{tabular}[c]{@{}c@{}}Evaluating the player’s healthy statement, \\ Predicting the player’s rehabilitation effect\end{tabular}                                                                & \begin{tabular}[c]{@{}c@{}}\cite{7995427}\cite{2017Cognitive}\cite{8007530}\cite{8834484}\\ \cite{2019Improving}\cite{ciman2018serious}\cite{anders2018exergames}\cite{8994565}\\ \cite{9069313}\cite{6289368}\cite{7319424}\end{tabular}                                            \\ \hline
Sensor                & \begin{tabular}[c]{@{}c@{}}Collecting player’s information of statement, \\ Sensor-based game controller, \\ Sensor-based assessment methods, \\ AR’s implementation, Haptic feedback\end{tabular} & \begin{tabular}[c]{@{}c@{}} \cite{8009394}\cite{7995427}\cite{8604049}\cite{ma2017comparison}\\ \cite{8834484}\cite{ciman2018serious}\cite{anders2018exergames}\cite{trombetta2017motion}\\ \cite{eisapour2020participatory}\cite{8882472}\cite{8088461}\cite{8001435}\\ \cite{6775492}\cite{8538058}\cite{7881304}\cite{6289368}\end{tabular} \\ \hline
AI and CI             & \begin{tabular}[c]{@{}c@{}}Predicting the player’s rehabilitation effect, \\ Difficulty adaptation, Virtual therapist, \\ Feedback control, Telerehabilitation\end{tabular}                        & \begin{tabular}[c]{@{}c@{}}\cite{6967825}\cite{8007530}\cite{6091665}\cite{9073659}\\ \cite{8001435}\cite{6775492}\end{tabular}                                                                                       \\ \hline
Telerehabilitation    & \begin{tabular}[c]{@{}c@{}}Home rehabilitation, Reducing medical expenses, \\ Using family factors to assist treatment, \\ Realizing exercise rehabilitation anytime and anywhere\end{tabular}     & \cite{6967825}\cite{9069313}\cite{8882432}\cite{7325501}                                                                                                                                                  \\ \hline
\end{tabular}
\end{table*}
~\\ 
~\\ 
\par \textbf{Technology development perspective}
\par $\bullet$SGERs mostly choose VR or AR as their game environment. The equipment they use is usually a screen or HMD. HMD has a large body, and some players will feel uncomfortable wearing it; the immersion provided by the screen is significantly lower than HMD. Therefore, we recommended that researchers choose projection AR devices when developing SGERs. The reasons were as follows: 1) The projection AR device can project the game into the real world without the player wearing an HMD; 2) the player can directly interact with the projected game, which can provide sufficient immersion and interaction.
\par $\bullet$To ensure the safety of the treatment process, the fixed difficulty SGER needs to be completed under the supervision of the therapist. This is because the excessive difficulty of the game will cause too much pressure on the player. The SGER that automatically adjusts the difficulty does not require the supervision of the therapist. The SGER monitors the player's performance in real time by sensors, reduces the difficulty when the player is tired and increases the difficulty when the player is eager for a challenge so that the player is always in a suitable state. At present, the automatic adjustment of game difficulty is accomplished by CI. Besides, the fixed difficulty SGER may waste time. The player may have obtained all the healing effects in half of the game, so the precious treatment time in the second half will be wasted. Therefore, we recommended that researchers design more reasonable, smarter, and more accurate difficulty adaptation methods when developing SGERs.
\par $\bullet$Feedback includes visual feedback, tactile feedback, and auditory feedback. It has two functions: reminding and guiding. Choosing an appropriate feedback method will greatly improve the player's sense of immersion and participation. So, we recommended that researchers invent more suitable feedback techniques for players. But in this process, we hoped that researchers can pay special attention to the feedback standards. How to give the player appropriate feedback at the right time is the key to success.
\par $\bullet$The accuracy of the assessment of the player's rehabilitation effect is very important because it means whether the treatment is effective and if it is effective, what is the effect. In order to obtain more accurate assessment results, we encouraged researchers to explore more assessment indicators (for example physiological signals) and assessment methods (such as the synthesis of multiple clinical scales, sensor-based assessment methods). Besides, predicting the player's effect of rehabilitation is also a direction we should pay attention to.
\par $\bullet$Sensor technology is widely used in SGER's exercise rehabilitation process. But there are still some problems. Such as, the volume of the sensor-based game controller is too big to the player's exercise, and there are fewer sensor-based assessment methods, and so on. Therefore, we recommended that researchers develop light-weight (preferably wear-free), inexpensive, and powerful sensor-based game control devices, develop more sensor-based assessment methods, and study the applicability of more types’ sensors to SGER.
\par $\bullet$To develop SGERs with higher simulation, safety, and usability, AI and CI are indispensable. AI and CI not only make the timing of feedback and difficulty adaptation more accurate but also make the image of the virtual therapist in the game more realistic and make its decision-making more scientific and reasonable. It can also collect and analyze the player's rehabilitation experience reasonably and efficiently to help later players perform better rehabilitation. Therefore, AI and CI are directions worthy of further research by researchers.
\par $\bullet$To reduce the inconvenience of patients going to the treatment site and promote the sharing of exercise rehabilitation experience, attention should be paid to the development of telerehabilitation. The rapid development of the Internet facilitates the real-time transmission of the player's game data to ensure that therapists can observe players almost simultaneously. Cloud storage facilitates the safe preservation of user data and the sharing of rehabilitation experience. The rise of AI and CI helps to achieve temporary replacement of therapists. When users want to do exercise rehabilitation and the therapist is not online, the virtual therapist created by AI can complete the therapist's task within the allowed range. Considering human creativity, virtual therapists cannot completely replace the therapists. When the virtual therapist and the therapist disagree, the therapist's opinion should be taken into account. 
~\\ 
\par \textbf{Functional design and social popularization perspective}
\par All types of SGERs:
\par $\bullet$When we reviewed SGER for integrated rehabilitation, we found in \cite{8994505} that certain multiplayer game modes are beneficial to players' exercise rehabilitation. Although we have not found games that use multiplayer mode in other types of SGER, we believed that multiplayer mode can be applied to all types of SGER. Because the multiplayer game mode affected the player's game experience and sharing of experience, which was not inconsistent with the type of exercise rehabilitation. Therefore, we believed that the multiplayer game mode is worthy of further research and use.
\par $\bullet$To provide researchers with a deeper understanding and help researchers develop better SGERs, further study serious games and exercise rehabilitation at the information and physical levels.
~\\ 
\par SGER for sub-healthy people:
\par $\bullet$Further analyze the sub-healthy people in society and find out the groups among them who need exercise rehabilitation.
\par $\bullet$Increase the promotion of SGER for sub-healthy people and encourage sub-healthy people to do exercise rehabilitation to avoid pathological changes in their bodies.
~\\ 
\par SGER for cognitive rehabilitation:
\par $\bullet$Further study the impact of different content on cognitive rehabilitation. For example, cartoon, dance, etc.
~\\ 
\par SGER for motor rehabilitation:
\par $\bullet$More traditional exercise rehabilitation programs can be programmed to improve the efficiency of SGERs.
\par $\bullet$Further study SGER for motor rehabilitation from the level of muscles, bones, and nerves.
\par $\bullet$The choice of the game controller should be noted. For example, robot exoskeleton-assisted exercises should be selected for those patients who have insufficient exercise capacity and cannot complete exercise rehabilitation.
~\\ 
\par SGER for integrated rehabilitation:
\par $\bullet$To help more patients who suffer from both cognitive impairment and dyskinesia, develop more SGERs with dual cognitive tasks. While developing SGERs, researchers should pay attention to the effects of different dual cognitive tasks on different diseases to design more targeted SGERs.
\par $\bullet$To help patients complete rehabilitation training, more consideration, and satisfaction of patients' psychological needs. Such as, providing more exercise assistance or cognitive assistance, providing more interesting game contents, etc.

\section{Conclusion}
Serious games have been widely studied as an effective way to achieve exercise rehabilitation. Although there were many studies, there was no clear definition of SGER. Therefore, we introduced exercise rehabilitation and SGER in detail to help researchers understand SGER more deeply. Then, we made a systematic summary of the development of SGER from the perspective of changes in the interactive mode based on the hardware. From this perspective, we divided the SGER into three stages: the stage of SGER's interaction by the game props, the stage of SGER’s interaction by the traditional controller, the stage of SGER’s interaction by the somatosensory controller, which can help researchers to understand the development of SGER efficiently and quickly. Besides, we proposed a SGER’s classification method based on degrees of unhealthiness different groups of people. Then, based on this classification method, we reviewed researches of SGER in the past decade years to help researchers better develop SGERs. Finally, we discussed current SGERs from technology development, functional design and social popularization perspectives and proposed some further research directions. To be brief, our work aimed to help researchers better understand the SGER and its development, and better develop more effective SGERs.


%



\section*{Acknowledgment}

This work was supported by the National Natural Science Foundation of China (61872038), and in part by the Fundamental Research Funds for the Central Universities under Grant FRF-GF-19-020B.

\ifCLASSOPTIONcaptionsoff
  \newpage
\fi



%


\bibliographystyle{IEEEtran}
\bibliography{A_Review_on_Serious_Games_for_Exercise_Rehabilitation}

%








\begin{IEEEbiography}[{\includegraphics[width=1in,height=1.25in,clip,keepaspectratio]{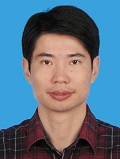}}]{Huansheng~Ning} Huansheng~Ning(M'10-SM'13) received the B.S. degree from Anhui University, Hefei, China, in 1996, and the Ph.D. degree from Beihang University, Beijing, China, in 2001.

He is currently a Professor and the Vice Dean with the School of Computer and Communication Engineering, University of Science and Technology Beijing, Beijing. He has authored 6 books and over 180 papers in journals and at international conferences/workshops. His current research interests include Internet of Things, Cyber Physical Social Systems, Cyberspace Data and Intelligence.

Prof. Ning is the founder and principal at Cybermatics and Cyberspace International Science and Technology Cooperation Base. He has been the Associate Editor of IEEE Systems Journal, the associate editor (2014-2018), area editor (2020-) and the Steering Committee Member of IEEE Internet of Things Journal (2018-). He is the host of the 2013 IEEE Cybrmatics Congress and 2015 IEEE Smart World Congress. His awards include the IEEE Computer Society Meritorious Service Award and the IEEE Computer Society Golden Core Member Award. In 2018, he was elected as IET Fellow.
\end{IEEEbiography}

\begin{IEEEbiography}[{\includegraphics[width=1in,height=1.25in,clip,keepaspectratio]{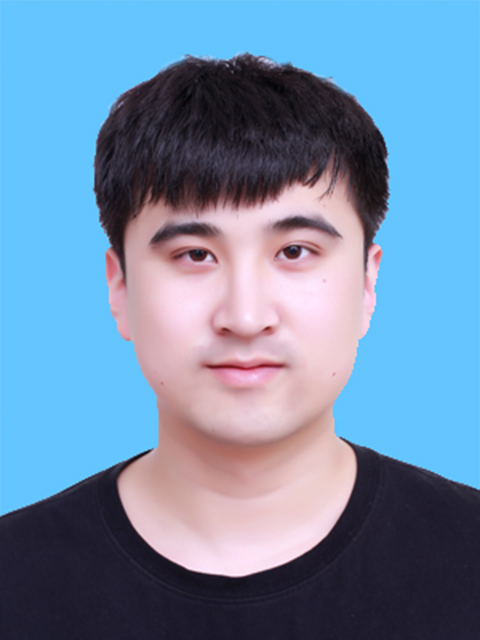}}]{Zhenyu~Wang} Zhenyu~Wang received the B.S. degree from Shandong Jianzhu University and currently working toward the M.S. degree in the School of Computer and Communication Engineering, University of Science and Technology Beijing, China.

His current research focuses the application of serious games.
\end{IEEEbiography}

\begin{IEEEbiography}[{\includegraphics[width=1in,height=1.25in,clip,keepaspectratio]{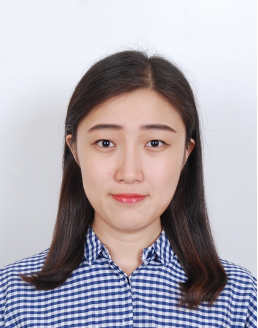}}]{Rongyang~Li} Rongyang~Li received the B.S. degree from Hebei Normal University of Science and Technology and currently working toward the M.S. degree in the School of Computer and Communication Engineering, University of Science and Technology Beijing, China.

Her current research focuses the user experience of games.

\end{IEEEbiography}

\begin{IEEEbiography}[{\includegraphics[width=1in,height=1.25in,clip,keepaspectratio]{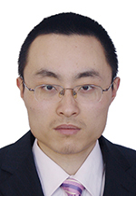}}]{YuDong~Zhang} Yu-Dong~Zhang received his PhD degree from Southeast University in 2010. He worked as a postdoc from 2010 to 2012 in Columbia University, USA, and as an assistant research sci- entist from 2012 to 2013 at Research Foundation of Mental Hygiene (RFMH), USA. He served as a full professor from 2013 to 2017 in Nanjing Normal University, where he was the director and founder of Advanced Medical Image Processing Group in NJNU. Now he serves as Professor in Department of Informatics, University of Leicester, UK.
\end{IEEEbiography}

\begin{IEEEbiography}[{\includegraphics[width=1in,height=1.25in,clip,keepaspectratio]{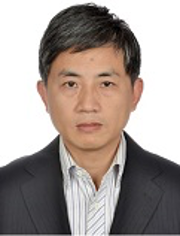}}]{Lingfeng~Mao} Lingfeng~Mao received the Ph.D. degree from Peking University, Beijing, China, in 2001.

He is currently a Professor with the School of Computer and Communication Engineering, University of Science and Technology Beijing, Beijing, and the Beijing Engineering Research Center for Cyberspace Data Analysis and Applications, Beijing. His current research interests include the modeling and characterization of semiconductors, semiconductor devices, integrated optic devices, and microwave devices.
\end{IEEEbiography}

\end{document}